\documentclass[12pt]{article}

 at 10truept

\def\rdots{\mathinner{\mkern1mu\raise1pt\vbox{\kern1pt\hbox{.}}\mkern2mu
   
\raise4pt\hbox{.}\mkern2mu\raise7pt\hbox{.}\mkern1mu}}
\newcommand{\be}{\begin{equation}}
\newcommand{\ee}{\end{equation}}
\newcommand{\ve}{\varepsilon}
\newcommand{\la}{\lambda}
\newcommand{\al}{\alpha}
\newcommand{\pa}{\partial}
\newcommand{\om}{\omega}
\newcommand{\Om}{\Omega}
\newcommand{\ga}{\gamma}
\newcommand{\ra}{\rightarrow}
\newcommand{\Z}{{\rm Z\kern-.35em Z}}
\newcommand{\bP}{{\rm I\kern-.15em P}}
\newcommand{\Q}{\kern.3em\rule{.07em}{.65em}\kern-.3em{\rm Q}}
\newcommand{\R}{{\rm I\kern-.15em R}}
\newcommand{\h}{{\rm I\kern-.15em H}}
\newcommand{\C}{\kern.3em\rule{.07em}{.55em}\kern-.3em{\rm C}}
\newcommand{\T}{{\rm T\kern-.35em T}}

\include{epsf}

\begin{document}    

\openup 1.5\jot
\vspace*{.7cm}
\centerline{\Large{The Einstein-Dirac-Maxwell Equations -- Black Hole
Solutions\footnote{To appear in the Proceedings of the IMS Conference on
Differential Equations in Mechanics, Chinese University of Hong Kong,
June 1999.}}}

\vspace{.5cm}
\centerline{by F.\ Finster\footnote{Max Planck Institute for 
Mathematics in the Sciences, Leipzig, Germany.},
 J.A.\ Smoller\footnote{University of Michigan, Ann Arbor, MI,  USA;
 research supported in part by the NSF, Grant No. DMS-G-9501128.},
 and S.-T.\ Yau\footnote{Harvard University, Cambridge, MA, USA; research
 supported in part by the NSF, Grant No. 33-585-7510-2-30.}}

\section{Introduction.}
\ \ \ \ \ We are interested in studying how different force fields 
interact with gravity, at a ``fundamental" level, but not 
at the level of Quantum Field Theory.  This is because there exists 
no theory of quantum gravity, and no understanding of ``Planck-scale" 
physics; that is, physics at ``Planck-lengths", where the Planck-length
is given by
\[	\left(  \frac {Gh}{c^3} \right)^{1/2} \approx 10^{-33} {\mbox{cm}}
\ ,	\]
where $G$ is Newton's gravitational constant, $h$ is Planck's constant,
and $c$ denotes the speed of light.

Before discussing our results, we think that it is worthwhile to see 
what is the difficulty in making a theory of quantum gravity.  In 
order to understand why a solution to the problem of reconciling 
gravity and quantum mechanics has been so elusive, we must consider 
the implications of the Heisenberg Uncertainty Principle (HUP) at 
small distance scales.

Recall that the HUP states that ``the more precisely a spatial 
measurement is made, the less precisely the momentum (or the energy) 
of the system being measured, is known".  When the spatial 
measurement $\Delta x \approx 10^{-13}$ cm, there are large 
uncertainties in the energy $E$.  These are realized as 
``fluctuations" at small distances.  Since $E = mc^2$, the energy of 
these fluctuations can give rise to ``virtual" particles and 
anti-particles, which arise out of the vacuum for short times before 
annihilating each other, and give rise to a ``sea of virtual 
particles".  When  
$\Delta x \approx 10^{-33}$ cm (Planck length), the energy 
fluctuations are quite large over small distances, and general 
relativistic effects become important.  In fact, from a theorem of 
Schoen and Yau [12], singularities must form, and these are believed 
to be black holes.  Thus  space-time becomes very curved at small 
distance scales; physicists say that space-time is ``foamy" (not 
smooth).  This invalidates the usual computational techniques of 
Quantum Field Theory, where small curvature is needed -- the 
calculations break down at high energies.  This suggests that at high 
energies, General Relativity, or Quantum Mechanics, (or both) must be 
modified.  The unsolved problem is, how is this to be done?

In order to set the background for our discussion, we shall briefly 
review what we consider to be some of the most important results concerning 
the coupling of gravity to other fields.  The first such result is 
due to Reissner, and Nordstr\"{o}m (1918, 1919) whereby they coupled 
gravity to electromagnetism, and this led to the celebrated 
Reissner-Nordstr\"{o}m solution, about which we shall have more to 
say in the next sections.  In the 1920's, Einstein, de Sitter, 
Friedmann, and others coupled gravity to perfect fluids, in order to 
study problems in Cosmology, the large scale structure of the 
Universe.  In the 1930's Oppenheimer, Tolman, Snyder, Volkoff, 
Landau, and others also coupled gravity to perfect fluids in order to 
study problems connected with stellar formation and collapse.  (All 
of the above results are classical and can be found in most standard 
textbooks on General Relativity; see e.g. [1].)

In the more modern era, the first important ideas stem from the 1988 
paper of R. Bartnik and J. McKinnon, [2].  These authors coupled 
Einstein's equations to an $SU(2)$ Yang-Mills field (that is, gravity 
to the weak nuclear force), and they found, numerically, 
particle-like solutions.  In the papers [15, 16], Smoller, 
Wasserman, and Yau rigorously proved the existence of both 
particle-like and black-hole solutions for the SU(2) 
Einstein/Yang-Mills equations.  Also in recent years, Smoller and 
Temple [13, 14]
coupled gravity to perfect fluids and they constructed a theory of 
shock-waves in General Relativity, with applications to Cosmology and 
stellar structure.  In the present paper, we couple gravity to both 
spinors and electromagnetism; i.e., we couple Einstein's equations to 
both the Dirac equation and Maxwell's equations and we investigate 
the existence of black-hole solutions; (in the works [4, 5], we consider
particle-like solutions of these equations).  We 
shall show here that in all cases considered, black-hole solutions do 
not exist; for complete details, see [6, 7], and also [8, 9] for related 
work.

\vspace{.25in}
\section{Background.}

\ \ \ \ \ In this section we shall give a short discussion of 
Einstein's equations of General Relativity, Maxwell's equations of 
electromagnetism, and Dirac's equation of relativistic Quantum 
Mechanics.  For more details, the reader should consult the standard 
textbooks, e.g. [1, 11]. 

We begin with General Relativity (GR).  The subject of GR is based on 
Einstein's three important hypotheses:

\begin{itemize}
\item [($E_1$)]  The gravitational field is described by the metric $g_{ij}$ in 
4-d space-time; $g_{ij}$ is assumed to be symmetric:  $g_{ij} = 
g_{ji}, \ \ i,j = 0,1,2,3.$
\item[($E_2$)] At each point, the $4 \times 4$ symmetric matrix 
$[g_{ij}]$ can be diagonalized as $g_{ij} =$ diag (1, -1, -1, -1).  
\item[($E_3$)]  The equations of GR should be independent of the 
choice of coordinate system.
\end{itemize}
The hypothesis $(E_1)$ is Einstein's great insight, whereby he 
``geometrizes" the gravitational field.  $(E_2)$ means that there 
should exist ``inertial frames" at each point, so that Special 
Relativity is included in GR, and $(E_3)$ implies that the 
gravitational field equations should be tensor equations.

The metric $g_{ij} = g_{ij}(x)$, $i,j =0,\ldots,3$, $x=(x^0, x^1, x^2, 
x^3)$, $x^0 = ct$, is the metric tensor defined on 4-d space-time.  {\it 
Einstein's equations} are ten (tensor) equations for the metric 
(gravitational field) and take the form
\be	R_{ij} - \frac 1 2 \: R \:g_{ij} \;=\; \sigma \:T_{ij} \ .	\ee
The left-hand side $G_{ij} \equiv R_{ij}- \frac 1 2 \: R \:g_{ij}$
is the {\it Einstein tensor}, and is a {\it geometric} object, while $T_{ij}$ the 
energy-momentum tensor, represents the source of the gravitational 
field and encodes the distribution of matter.  The word ``matter" in 
general relativity refers to everything which can produce a 
gravitational field.  The tensor $T_{ij} $ is required to satisfy 
the relation $T^i_{j;i} = 0$, (it's covariant divergence vanishes), 
and this in turn expresses the laws of conservation of energy and 
momentum.  The quantities which comprise the Einstein tensor $G_{ij}$ 
are given as follows:  First from the metric tensor $g_{ij}$, we form 
the {\it Levi-Civita connection} $\Gamma^k_{i,j}$:
\[	\Gamma^k_{ij} = \frac 1 2 \; g^{k\ell} \left( \frac{\pa g_{\ell 
j}} {\pa x^i} + 
            \frac{\pa g_{i \ell}} {\pa x^i} - \frac{\pa g_{i j}} {\pa 
x^\ell} \right) , \]
where $[g^{k\ell}] = [g_{k\ell}]^{-1}$, and summation convention is 
employed; namely an index which appears as both a subscript and 
superscript is to be summed form 0 to 3.  Having $\Gamma^k_{ij}$, we 
then construct the {\it Riemann curvature tensor} $R^i_{qk\ell}$:
\[	R^i_{qk\ell} = \frac {\pa\Gamma^i_{q \ell}} {\pa x^k} - \frac 
{\pa\Gamma^i_{qk}} {\pa x^\ell} + \Gamma^i_{pk} \Gamma^p_{q\ell} - 
\Gamma^i_{p\ell} \Gamma^p_{qk}.  \]
Finally, we can explain the terms $R_{ij}$ and $R$ in $E_{ij}$; 
namely $R_{ij} = R^s_{isj}$ is the {\it Ricci tensor}, and the scalar 
$R = g^{ij} R_{ij}$, is called the {\it scalar curvature}.

The quantity $\sigma$ is a universal constant defined by
\[	\sigma = \frac{8\pi G}{c^4},	\]
where $G$ is the gravitational constant, and $c$ is the speed of light.

From these definitions, one immediately sees the enormous complexity 
of the equations (2.1) for the unknown quantities $g_{ij}$.  For this 
reason, one seeks solutions which are highly symmetric, and in what 
follows, we shall only consider solutions which are spherically 
symmetric; i.e. the metric has the form
\be	ds^2 = T^{-2}(r) \:dt^2 - A^{-1}(r) \:dr^2 - r^2 \:d\Omega^2,	\ee
where $d\Om^2 = d\vartheta^2 + \sin^2 \vartheta \:\varphi^2$ is the standard 
metric on the 2-sphere, $(r, \vartheta, \varphi)$ are the usual polar
coordinates, and $t$ denotes time.  Such metrics turn out to be quite 
interesting mathematically, in addition to having physical interest.

Now consider the problem of finding the gravitational field exterior 
to a ball in $\R^3$; that is, there is no matter exterior to the 
ball.  This actually models well our solar system, where we think of 
the Sun as the 3-ball, and we ignore the masses of the planets.  In 
this case the Einstein equations become $R_{ij} - \frac 1 2 \: 
R \:g_{ij} = 0$, and we seek spherically symmetric solutions.  This 
problem was solved by K. Schwarzschild in 1916, and its solution is 
called the {\it Schwarzschild metric}:
\[	ds^2 = \left( 1 - {2m \over r} \right) c^2dt^2 - \left( 1 - {2m 
\over r} \right)^{-1} dr^2 
                     - d\Om^2.	\]
Here $m = GM/c^2$ and $M$ is the mass of the 3-ball.  Since $2m$ has 
dimension of length, it is called the Schwarzschild radius.  One sees 
immediately that the Schwarzschild metric is singular at $r=2m$; 
namely $g_{00} = 0$ and $g_{11} = \infty$.  For $r < 2m$, the signs 
of the metric components change:  $g_{00} >0$ and $g_{11} < 0$.  So 
we must reconsider the physical meaning of $t$ and $r$ inside the 
Schwarzschild radius.

\begin{figure}[tb]
        \centerline{\epsfbox{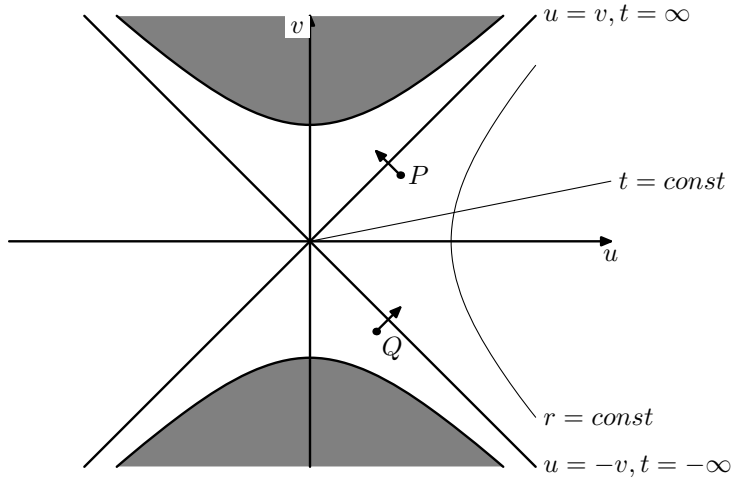}}
        \caption{Kruskal Coordinates}
        \label{fig1}
\end{figure}

To investigate the mathematical and physical nature of the 
Schwarzschild radius, we introduce Kruskal coordinates $(r,t) 
\longrightarrow (u,v)$, and we seek a transformed metric of the form
\[	ds^2 = f(u,v) (dv^2 - du^2) +r^2 \:d \Om^2 .		\]
This yields the following in the region $r > 2m$,  (see Figure 
\ref{fig1})
\begin{eqnarray*}
u &=& \sqrt{ \frac r{2m} -1} \ e^{r/4m} \cosh \frac {ct}{4m} \\
v &=& \sqrt{ \frac r{2m} -1} \ e^{r/4m} \sinh \frac {ct}{4m} \\
f &=&  \frac {32m^3}{r}  \ e^{r/4m} ,
\end{eqnarray*}
with similar expressions in the region $r < 2m$.

We can use Kruskal coordinates to study light rays traveling radially 
inward towards the Schwarzschild radius, starting at a point $P$ 
outside of the Schwarzschild radius; i.e., in the region $r > 2m$; 
see Figure 1.  In $u/v$ coordinates, the light ray leaves at the 
point $P$ in $r > 2m$, $t$ finite, and travels inward towards the 
Schwarzschild radius $r=2m$, as $t \rightarrow \infty$.  It crosses 
the line $t=\infty$ into the interior of the Schwarzschild sphere.  
Incoming light is thus in effect, totally absorbed by the 
Schwarzschild sphere. 

We can also study light emitted from inside the Schwarzschild sphere 
starting at a point $Q$ in the region $r < 2m$; again see Figure 1.  
The trajectory starts at some $r < 2m$ and finite $t$, travels 
through increasing $r$ but decreasing $t$, and crosses the 
Schwarzschild radius $t = -\infty$ to the exterior of the 
Schwarzschild sphere, where its evolution is normal.  Thus, light 
emerging from the Schwarzschild sphere must have been traveling since 
before $t = -\infty$; in effect since before the beginning of time.  
It is questionable whether such light is physically observable.  If 
not, the Schwarzschild sphere has the physical properties of a 
black-hole:  it absorbs all light and emits none.

In general, for a metric of the form (2.2), we define a {\it 
black-hole solution} to be a solution of Einstein's equations which 
satisfies
\be	A(\rho) = 0, \qquad {\rm and} \qquad A(r) > 0 \ \ \ {\rm if} \ \ 
\ r > \rho ,	\ee
for some $\rho > 0$; $\rho$ is the radius of the black-hole and is 
referred to as the ``event horizon".

\begin{figure}[tb]
        \centerline{\epsfbox{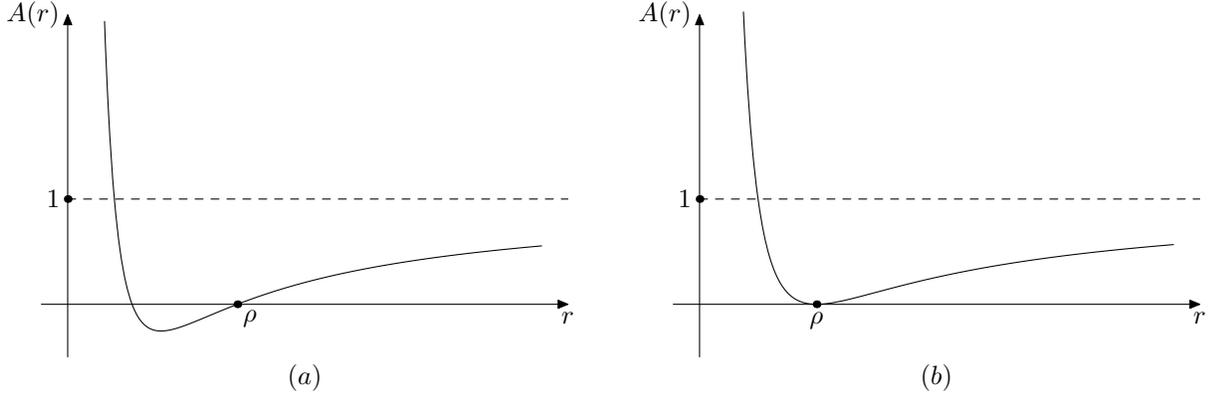}}
        \caption{The Reissner-Nordstr\"om Black Hole}
        \label{fig2}
\end{figure}

We now consider, in a coordinate invariant way, Maxwell's equations for an 
electromagnetic field.  First we let\footnote{Throughout this paper 
we use summation convention.}

\[	{\cal A} \;=\; A_i \:dx^i	\]
be a 1-form, called the electromagnetic potential (connection), and 
the 2-form $F=dA$ is the associated electromagnetic field; in 
coordinates
\[  F = F_{ij} \:dx^i \wedge dx^j \;, \ \ \ \ \ F_{ij} = {\pa A_j \over 
\pa x^i} - {\pa A_i \over \pa x^j} \ .   \]
Maxwell's energy momentum tensor is
\[
T_{ij} = {1 \over 4\pi} \left[ g^{\ell m} F_{i\ell} F_{jm} - \frac 1 
4 F_{\ell m} F^{\ell m} g_{ij}\right],
\]
where as usual, we always use the metric to raise the indices; thus 
$F^{\ell m} = g^{\ell i} g^{mj} F_{ij}$.  Maxwell's equations are the 
pair of equations
\begin{eqnarray*}
dF &=& 0 \ \ \ \ \ \ \ \ \ \ {\rm (automatic)} \\
d^*F &=& 0 \ \ ,
\end{eqnarray*}
where ``$^*$" denotes the Hodge star operator, mapping 2-forms 
into themselves, defined by
\[	(^* F)_{k\ell} = \frac 1 2 \ \sqrt{|g|} \:\ve_{ijk\ell} \:F^{ij} \ , \]
where $g = \det (g_{ij})$, and $\ve_{ijk\ell}$ is the completely 
anti-symmetric symbol defined as $\ve_{ijk\ell} =$ sgn$(i, j, k, 
\ell)$.  It is important to notice that the *-operator depends on the 
metric ($g_{ij}$).  The question we now ask is the following:  Find 
the solution of the Einstein/Maxwell equations outside of a {\it 
charged} ball in $\R^3$.  That is, we seek the spherically symmetric 
static, exterior (gravitational and electromagnetic) field of a 
charged distribution of matter.  This problem was solved in 1918-1920 
by Reissner and Nordstr\"{o}m, and the gravitational field is 
described by the metric
\[	ds^2 = \left( 1 - {2m \over r} + {q \over r^2} \right)
	dt^2 - \left( 1 - {2m \over r} + {q \over r^2} \right)^{-1} dr^2 - 
r^2 \:d\Om^2 ,    \]
where $m = GM/c^2, q = 2\pi GQ^2/c^4$; $M$ and $Q$ denote the mass 
and charge respectively, of the ball.  Notice that if $Q$ is 
sufficiently large, the metric coefficient $A(r) = 1 - {2m \over r} + 
{q \over r^2} $ is never zero; otherwise there are two possibilities 
(see Figure \ref{fig2}).
Case (a) is called a non-extreme Reissner-Nordstr\"{o}m solution and Case (b) is 
called an extreme Reissner-Nordstr\"{o}m solution; these black-hole 
solutions will come up again in the next sections.

We now turn to the Dirac equation.  The Dirac equation brings in 
Quantum Mechanics and particles into our work, and it also describes 
the intrinsic ``spin" of the particle (fermion).  The Dirac equation 
can be written as
\be	(G-m) \:\Psi = 0		\ee
where $G$ is the Dirac operator, and $\Psi$ is the wave function of a 
fermion (proton, anti-proton, electron, positron, neutrino, etc.) 
having (rest) mass $m$.  $\Psi$ is a complex 4-vector and is called a 
{\it spinor}.  The Dirac operator $G$ takes the form
\[	G = iG^j(x) \ {\pa \over \pa x^j} + B(x),	\]
where the $4 \times 4$ matrices $G^j$ are called Dirac matrices, and 
$B$ is a $4 \times 4$ matrix.  The Dirac matrices and the Lorentzian 
metric $g^{jk}$ are related by the anti-commutation relations
\be	g^{jk} I = \frac 1 2 \{ G^j, G^k \} = \frac 1 2 (G^j G^k + G^k 
G^j).	\ee
Now let $\cal H$ be a space-like hypersurface with a future directed 
normal vector field $\nu$, and define an inner product on solutions 
of the Dirac equation by 
\[	(\Phi, \Psi) = \int_{\cal H} \overline{\Phi} G^i \Psi \:\nu_j \:
d\mu \ ,	\]
where
\[	\overline{\Phi} = \Phi^* \left(
\begin{array}{cc}
I & 0 \\
0 & -I
\end{array} \right)
\]
is called the {\it adjoint spinor},  * denotes complex conjugation, 
and $\mu$ is the invariant measure on $\cal H$ induced by the 
metric.  This inner product is positive definite and independent of 
$\cal H$ because of {\it current conservation},
\[	\nabla_i \overline{\Phi} G^i \Psi = 0.	\]
From a direct generalization of $\overline{\Psi} \gamma^0 \Psi$ in 
$M^4_1$ (Minkowski space), where $ \gamma^0 = \left( \begin{array}{cc}
I & 0 \\
0 & -I
\end{array} \right)$, $\overline{\Psi} G^j \Psi \nu_j$ is interpreted as a 
{\it probability density} and for non-black-hole solutions, we 
normalize solutions of the Dirac equation by the requirement
\[	(\Psi, \Psi ) = 1.		\]
For a black-hole solution with event horizon $\rho$, we demand that 
(cf.~[6,7]), 
\be 	\int_{\{t={\rm const.}, r>r_0\}} \overline{\Psi} G^j \Psi \nu_j \;
d\mu \;<\; \infty, \ \ {\rm for \; all} \ \ r_0 > \rho .	\ee

Now a result in [3] allows us to choose the Dirac matrices $G^j$ to 
be any $4 \times 4$ matrices which are Hermitian with respect to the 
inner product
\[	\big< \Phi, \Psi \big> = \int_{\R^4} \overline{\Phi} \Psi \sqrt{|g|} 
d^4x \ ,		\]
and satisfy (2.5).  Here $B(x)$ is defined by
\[	B(x) = G^j(x) E_j(x) + G^j(x) A_j(x),		\]
where $A_idx^i$ is the em potential,
\[  E_j = \frac i 2 \; \rho(\pa_j\rho) - {i \over 16}\; Tr (G^m 
\nabla_j G^n) \:G_m G_n + \frac i 8 \; Tr(\rho G_j \nabla_m G^m) \rho
\]
is the spin connection, and
\[	\rho = \frac i 4 \:\sqrt{|g|}\; \ve_{ijk\ell} G^iG^jG^kG^\ell .	\]

In this framework, the {\it Einstein-Dirac-Maxwell} (EDM) equations 
for one particle are
\begin{eqnarray*}
R_{ij} - \frac 1 2 \; Rg_{ij} &=& \sigma T_{ij} \\
(G-m) \Psi &=& 0 \\
\nabla_k F^{jk} &=& 4\pi e \:\overline{\Psi} G^j \Psi \;\;\ ,
\end{eqnarray*}
where $T_{ij}$ is the Dirac energy-momentum tensor, and
where $e$ denotes the charge of the Dirac particles.  These are 18 
PDE's for the 18 unknowns $g_{ij}(10), A_i(4)$, and $\Psi(4)$; where 
$\Psi$ is complex.  In this generality, the equations are hopelessly 
complicated, and are impossible to analyze.  We thus specialize by 
imposing certain symmetry conditions; see [5, 6] for complete details.

For $j = \frac 1 2, \frac 3 2, ....$ we consider a static, 
spherically symmetric system of $(2j+1)$ Dirac particles, each having 
angular momentum $j$.  (In the language of atomic physics, we 
consider the completely filled shell of states with angular momentum 
$j$.  Classically, one can think of this multi-particle system as 
several Dirac particles rotating around a common center 
such that their total angular momentum is zero.)  Since the system of 
fermions is spherically symmetric, we can obtain a consistent set of 
equations by assuming that both the gravitational and em fields are 
spherically symmetric.  This allows us to separate out the angular 
dependence and reduce the problem to a system of nonlinear ODE's.  
Thus taking the metric in the form (2.2) and the em potential in the 
Coulomb gauge, ${\cal{A}}=\phi(r) \:dt$, we can take for the Dirac matrices the 
following:
\begin{eqnarray*}
G^t &=& T\gamma^0 \\
G^r &=& \sqrt{A} \big( \ga^1 \cos \vartheta + \ga^2 \sin \vartheta \cos 
{\varphi} + \ga^3 \sin \vartheta \sin {\varphi} \big) \\
G^\vartheta &=& \frac 1 r \big(- \ga^1 \sin \vartheta + \ga^2 \cos \vartheta 
\cos {\varphi} + \ga^3 \sin \vartheta \sin {\varphi} \big) \\
G^{{\varphi}} &=& \frac 1 {r \sin \vartheta} \big( - \ga^2  \sin 
{\varphi} + \ga^3 \cos {\varphi} \big) 
\end{eqnarray*}
where
\[	\ga^0 = \left( \begin{array}{cc}
I & 0 \\
0 & -I  \end{array} \right), \ \ \ 
\ga^0 = \left( \begin{array}{cc}
0 & \sigma^i \\
-\sigma^i & 0 \end{array} \right), \ \ \   i = 1,2,3 ,   \]
and $\sigma^1, \sigma^2, \sigma^3$ are the Pauli matrices (the 
standard basis for $su(2)$).

By using a suitable ansatz and taking the wave functions in the form 
$\Psi = e^{i\om t} f(r)$, we can reduce the complex 4-spinors to real 
two spinors
\[	\Phi \;=\; (\al, \beta), \ \ \ \ \al, \beta \ \ \ {\rm real}.	\]
The Dirac equations become
\begin{eqnarray}
\sqrt{A} \:\al ' &=& {2j+1 \over 2r} \ \al - [(\om - e\phi)T+m]\beta \\
\sqrt{A} \:\beta ' &=& [(\om - e\phi-m)T-m] \; \al - {2j+1 \over 2r} 
\beta \; ,
\end{eqnarray}
and the normalization condition (2.6) is the requirement that 
\be	\int^\infty_{r_0} (\al^2 + \beta^2) \ \frac {\sqrt{T}} A \; dr
\;<\; \infty,	\ee
for every $r_0 > \rho$ where $r=\rho$ defines the event horizon.

The full EDM equations consist of (2.7), (2.8), together with the 
Einstein equations
\begin{eqnarray}
rA' &=& 1 - A - 2(2j+1)(\om - e\phi) \:T^2\: (\al^2+\beta^2) - r^2 
AT^2(\phi ')^2 \\
2rA {T' \over T} &=& 1 - A - 2(2j+1)(\om - e\phi) \:T^2\: (\al^2+\beta^2) 
+ 2 \; {(2j+1)^2 \over r} \:T\: \al \beta \\
&+& 2(2j+1)\:mT\: (\al^2+\beta^2) + r^2 AT^2 \:(\phi ')^2 , \nonumber
\end{eqnarray}
and Maxwell's equation
\be	r^2 A \phi '' \;=\; -(2j+1) \:e\:(\al^2 + \beta^2) -
\left( 2rA + r^2A \:{T' \over T} + {r^2 \over 2} \:A' \right) \phi ' .	\ee
Notice that the EDM equations are invariant under the gauge 
transformations
\[	\phi \rightarrow \phi + k, \ \ \om \ra \om + ek, \ \ \ \ k \in 
\R.		\]
Finally, in addition to (2.9), it is required that the following 
conditions hold:
\be	\lim_{r \ra \infty} r \:(1 - A(r)) < \infty \ ,	\ee
(finite total mass),
\be	\lim_{r \ra \infty} T(r) = 1	\ee
(asymptotic flatness),
\be	\lim_{r \ra \infty} \phi(r) = 0 	\ee
(em potential vanishes at infinity).

We make the following 3 assumptions on the regularity properties of 
the event horizon $r=\rho$:

\begin{itemize}
\item[(I)]  The volume element $\sqrt{\det|g|}$ is smooth and 
non-zero on the horizon: $T^{-2} A^{-1}$ and $T^2A \in 
C^\infty([\rho, \infty))$.
\item[(II)] By definition, the strength of the em field is the scalar 
$F_{ij}F^{ij} = -2(\phi ')^2 AT^2$.  We assume this to be bounded 
near $r=\rho$, so that in view of (I), we have
\[	|\phi '(r)| < c_1, \ \ \ \ \rho < r < \rho + \ve_1 .   \]
\item[(III)]  $A(r)$ satisfies a power law near $r=\rho$,
\[	A(r) = c\:(r-\rho)^s + O(r-\rho)^{s+1}, \qquad s > 0.	\]
\end{itemize}
We remark that if (I) or (II) were violated, then an observer freely 
falling into the black-hole would encounter strong forces when 
crossing the event horizon; such as an effect is seen to be false 
when on passes to Kruskal coordinates.  Hypothesis (III) is a 
technical condition which includes all known physically relevant 
horizons; $s=1$ for Schwarzschild and Reissner-Nordstr\"{o}m 
black-holes, and $s=2$ for the extreme Reissner-Nordstr\"{o}m 
black-hole.  This condition could probably be weakened.

Here is our main result.

\noindent
\underline{Theorem}. {\it Every black-hole solution of the EDM 
equations satisfying Assumptions (I) - (III) is either}

a)  {\it a non-extreme Reissner-Nordstr\"{o}m solutions with $\al(r) 
\equiv 0 \equiv \beta(r)$, for all  $ r \ge \rho$, or}

b)  $s=2$ {\it and we find numerically that either the normalization 
condition (2.9) fails, or the solution is not regular everywhere 
outside the event horizon.}

Thus the only black-hole solutions of the EDM equations are 
Reissner-Nordstr\"{o}m solutions:  the spinors vanish identically, 
and so are not normalizable.  Thus these ``quantum effects" dis-allow 
(stationary) black-holes.  The result indicates too that the Dirac 
particles must either disappear inside the event horizon, 
or tend to infinity; namely there is zero probability for the 
particles to remain in a finite region outside of the black hole.

We now give some ideas of the proof; the reader should consult [6] 
for the full details.  We assume that the EDM equations admit a 
solution outside of the even horizon $r=\rho$, with $\big( \al(r), 
\beta(r) \big)  \not\equiv (0,0)$, and we shall obtain a 
contradiction.  We begin with the case $s < 2$.

\noindent
\underline{Lemma 1}.  {\it If $s < 2$, then there exist constants $c 
> 0$ and $\ve > 0$ such that}
\be	c \;\le\; \al^2(r) + \beta^2(r) \;\le\; \frac 1 c, \qquad
\rho < r < \rho +\ve.		\ee

\noindent
\underline{Proof}.  From the Dirac equations (2.7), (2.8) we have
\begin{eqnarray*}
\sqrt{A} \:(\al^2 + \beta^2)' &=& 2(\al, \beta)
\left( \begin{array}{cc}
{2j+1 \over 2r} & -m \\
 \\
-m & {2j+1 \over 2r}  
\end{array} \right)
\left( \begin{array}{c}
 \al \\ 
 \\
\beta \end{array} \right) \;\leq\;
\sqrt{ 4m^2 + {(2j+1)^2 \over r^2} }\: (\al^2 + 
\beta^2) \;\;\;.
\end{eqnarray*}
By uniqueness, $(\al^2 + \beta^2)(r) > 0$ on $\rho < r < \rho + \ve$ 
for any $\ve > 0$.  Dividing by $\sqrt{A} (\al^2 + \beta^2)$, 
integrating from $r > \rho$ to $\rho + \ve$, and noting that $A^{-\; 
1/2}$ is integrable near $\rho$ gives
\begin{eqnarray*}
\left| \log {(\al^2 + \beta^2)(\rho +\ve) \over (\al^2 + \beta^2)(r)} 
\right| &\le&
 \int^{\rho + \ve}_r \left(4m^2 + {(2j+1)^2 \over t^2} \right)^{1/2} 
A^{-\;1/2} (t)dt \\
&\le&  \int^{\rho + \ve}_\rho \left(4m^2 + {(2j+1)^2 \over t^2} 
\right)^{1/2} A^{-\;1/2} (t)dt < \infty,
\end{eqnarray*}
which implies the result.

\bigskip

\noindent
\underline{Proposition 2}.  $s < 2$ {\it cannot hold.}

\noindent
\underline{Proof}. From Einstein's equations (2.10), (2.11), we can 
write

$\ \ \ \ \ \ \ \  r(AT^2)' \;=\; -4(2j+1)(\om-e\phi) (\al^2 + \beta^2)T^4 
$

\[  +\left[ 2 {(2j+1)^2 \over r} \al \beta + 2(2j+1) m  (\al^2 - 
\beta^2) \right] T^3.
\]
By (I), the lhs is bounded, and $T \ra \infty$ as $r \searrow \rho$.  
Thus, in view of (2.16), we must have
\be 	\lim_{r \searrow \rho} \:(\om - e\phi (r)) = 0.		\ee
We next write Maxwell's equation (2.12) in the form
\be	\phi '' = - \frac 1 A {(2j+1)e \over r^2} \:[\al^2 + \beta^2] - 
\left( \frac 1 {r^2t\sqrt{A}} \:[r^2T \sqrt{A}]'\: \phi' \right).		\ee
Since the term ( \ \ \ ) is bounded near $r=\rho$, and $s \ge 1$ 
implies $A^{-1}$ is not integrable, we would conclude if $s \ge 1$ 
that $\phi '$ is unbounded, thereby violating (II).  It follows that 
we must have $s < 1$.

Now if we integrate (2.18), we obtain, for $r$ near $\rho$, 
\[	\phi '(r) \;=\; c_1\:(r - \rho)^{-s+1} + c_2 + O((r-\rho)^{-s+2}) .	\]
Integrating again, and using (2.17) gives
\[	\phi(r) \;=\; c_1(r - \rho)^{-s+2} + c_2 (r - \rho)+ \frac \om e + 
O((r-\rho)^{-s+3})\;\;\; , \]
and thus $(\om - e\phi) = O(r-\rho)$.  Now from (2.10)
\begin{eqnarray*} 
rA' &=& (1-A) - 2(2j+1)(\om-e\phi) \big[T^2(\al^2 + \beta^2)\big] - 
\big[r^2(AT^2)(\phi ')^2\big] \\
&=& O(1) - O((r-\rho)^{1-s}) - O(1) 
\end{eqnarray*}
so that the rhs is bounded, but $s < 1$ implies that the lhs blows 
up.  This contradiction shows that $s < 2$ cannot hold.

\medskip
\noindent
\underline{Proposition 3}.  $s > 2$ {\it cannot hold.}

\noindent
\underline{Proof}. In [6], we prove the following estimates
\be	\lim_{r\searrow \rho} (r-\rho)^{- \frac{s}{2}} \big(\al^2(r) + 
\beta^2(r) \big) \;=\; 0,   \ee
and
\be	\lim_{r\searrow \rho}  \phi '(r)^2 \;=\; \frac 1 {\rho^2} 
\lim_{r\searrow \rho} A^{-1}(r) T^{-2}(r) \;>\; 0.	\ee
Then (2.19) shows that near $r=\rho$, 
\[	(\om - e\phi)(r) = c+d\:(r-\rho) + o(r-\rho),  \ \ \ \ \ d\not= 0,	\]
and
\[   [(\om - e\phi)T]' = (r-\rho)^{-s}\:[d+c(r-\rho)^{-1}],		\]
so that as $r \searrow \rho$,
\[   [(\om - e\phi)T]	\longrightarrow \infty  \ \ \ \ {\rm 
monotonically}.  \]
Now the Dirac equations (2.7), (2.8) can be written as
\begin{eqnarray*}
\sqrt{A} \left( \begin{array}{c} 
\al \\
  \\ 
\beta 
\end{array} \right)' &=& 
\left(
\begin{array}{cc}
{ 2j+1 \over 2r} & -\left[(\omega - e\phi )T + m\right]  \\
 \\
  \left[(\omega - e\phi)T - m\right]  & {-2j+1 \over 2r} 
\end{array} 
\right)
  \left( \begin{array}{c}
 \al \\ 
 \\ 
\beta \end{array} \right) \\
&\approx & \left( \begin{array}{cc} 
a & -b \\
 \\
b & -a 
\end{array} \right) 
\left( \begin{array}{c}
 \al \\ 
 \\ 
\beta \end{array} \right),
\end{eqnarray*}
for $r$ near $\rho$, where $b \nearrow \infty$.  The eigenvalues of 
this last matrix satisfy $\la^2 \approx a^2 - b^2 \ra -\infty$ as $r 
\searrow \rho$.  This suggests that the vector $(\al, \beta)$ spins 
around the origin faster and faster as $r \searrow \rho$, and that 
$(\al, \beta)(r) \not\ra (0,0).$  In fact, we prove that
\[    \underline{\lim}_{r\searrow \rho} \left(\al^2(r) + \beta^2(r) 
\right) > 0 ,    \]
but this contradicts (2.19).  Thus Proposition 3 holds.
	
	The remaining case is $s=2$.  In this case the metric and electric 
field behave near the horizon like the extreme Reissner-Nordstr\"{o}m 
solution.  Physically, this means that the electric charge of the 
black-hole is so large that the electric repulsion balances the 
gravitational attraction and prevents the Dirac particles from 
entering the black hole.  Obviously, this is not the physical 
situation which one can expect, say, from the gravitational collapse 
of a star.  On the other hand, extreme Reissner-Nordstr\"{o}m 
black-holes have zero temperature [10, 17], and can be considered as 
end-states of black-holes emitting Hawking radiation.  It is thus 
interesting to see whether the case $s=2$ can admit normalizable 
solutions of the EDM equations.  Considering the Dirac equation in an 
extreme Reissner-Nordstr\"{o}m black-hole background field, it was 
proved in [6] that the normalization 
condition (2.6) is violated.  What we want to know here is whether 
the influence of the spinors on the gravitational and em fields can 
make the normalization condition finite.  Clearly this is a very
difficult mathematical problem.  We have made numerical 
investigations, and from these we conclude that the solutions of the 
EDM equations either develop singularities at some finite $r$, or the 
normalization condition fails.  Thus in the case $s=2$, our numerical 
investigations show that there are no normalizable solutions of the 
EDM equations.

Our results show that taking quantization and spin into account 
implies a breakdown of the classical situation; namely, there cannot 
exist normalizable black-hole solutions of the EDM equations.  Applied 
to the gravitational collapse of a ``cloud" of spin-1/2 -particles to 
a black hole, our result indicates that the Dirac particles must 
either disappear into the black-hole, or escape to infinity.

We remark that these results have been extended to the axisymmetric 
case including the Kerr-Newman rotating black hole [9]. In addition,
we have recently proved that there are no normalizable 
solutions of the  Einstein-Dirac-$SU(2)$ Yang/Mills equations; this 
result holds under a more general condition than (III), and applies 
for every $s > 0$.

We close with two facts.  First, our results are basically a 
consequence of the Heisenberg Uncertainty Principle, together with 
the specific form of the Dirac current.  Second, it is essential for 
our results that the particles have spin; spin is an important effect 
which must be considered in studying gravitational collapse.

\vfill\eject

\centerline{REFERENCES}

\begin{enumerate}
\item Adler, R., Bazin, M., and Schiffer, M., ``Introduction to 
General Relativity", 2nd ed. New York: McGraw-Hill (1975)
\item Bartnik. R., and McKinnon, J., ``Particle-like solutions of the 
Einstein-Yang-Mills equations", {\it Phys.\ Rev.\ Lett.}\ 61 (1988) 141-144
\item Finster, F., ``Local $U(2,2)$ symmetry in relativistic quantum 
mechanics", hep-th/9703083, {\it J.\ Math.\ Phys.}\ 39 (1998) 6276-6290
\item Finster, F., Smoller, J., Yau. S.-T., ``Particle-like solutions 
of the Einstein-Dirac equations", gr-qc/9801079, {\it Phys.\ Rev.\ 
D}\ 59 (1999) 104020
\item Finster, F., Smoller, J., Yau. S.-T., ``Particle-like solutions 
of the Einstein-Dirac-Maxwell equations", gr-qc/9802012,
{\it Phys.\ Lett.}\ A 259 (1999) 431-436
\item Finster, F., Smoller, J. , and Yau, S.-T. ``Non-existence of 
black hole solutions for a spherically  symmetric, static 
Einstein-Dirac-Maxwell system, gr-qc/9810048,
{\it{Commun.\ Math.\ Phys.}}\ 205 (1999) 249-262
\item Finster, F., Smoller, J. , and Yau, S.-T. ``Non-existence of 
time-periodic solutions of the Dirac equation in  a 
Reissner-Nordstr\"{o}m black hole background", gr-qc/9805050,
{\it{J.\ Math.\ Phys.}}\ (to appear)
\item Finster, F., Smoller, J. and Yau, S.-T., ``The interaction of 
Dirac particles with non-abelian gauge fields and gravity -- black 
holes,'' in preparation
\item  Finster, F., Kamran, N., Smoller, J., and S.-T. Yau, 
``Non-existence of time-periodic solutions of the Dirac equation in 
an axisymmetric black-hole geometry", gr-qc/9905047 (1999)
\item Hawking, S., ``Particle creation by black-holes, {\it Commun.\ 
Math.\ Phys.}\ 43 (1975) 199-220
\item Sakurai, J.J., Advanced Quantum Mechanics, Addison-Wesley (1967)
\item Schoen, R., and Yau, S.-T., ``The existence of a black hole due 
to condensation of matter", {\it Commun.\ Math.\ Phys.}\ 90 (1983)
575-579
\item Smoller, J., and Temple, B., ``Astrophysical shock-wave 
solutions of the Einstein equations", {\it Phys.\ Rev.}\ D 51 (1995)
2733-2743
\item Smoller, J., and Temple, B., ``Cosmology with a shock-wave",
astro-ph/9812063, {\it Commun.\ Math.\ Phys.}\ (to appear) 
\item Smoller, J., and Wasserman, A., ``Existence of infinitely-many 
smooth, static global solutions of the Einstein-Yang/Mills equations, 
{\it Commun.\ Math.\ Phys.}\ 51 (1993) 303-325
\item Smoller, J., Wasserman, A., and Yau, S.-T., ``Existence of 
black hole solutions for the Einstein-Yang/Mills equations", {\it  
Commun.\ Math.\ Phys.}\ 154 (1993) 377-401
\item Smoller, J. and Wasserman, A., ``Uniqueness of extreme 
Reissner-Nordstr\"{o}m solution in SU(2) Einstein-Yang/Mills theory 
for spherically symmetric space-time", {\it  Phys.\ Rev.}\ D 52 (1995) 
5812-5815
\end{enumerate}

\bigskip

\noindent
Felix Finster \\
Max Planck Institute for Mathematics in the Sciences \\
\noindent
Inselstr.\ 22-26\\
\noindent
04103 Leipzig, GERMANY \\
\noindent
Felix.Finster@mis.mpg.de

\medskip

\noindent
Joel Smoller \\
University of Michigan \\
\noindent
Mathematics Department \\
\noindent
Ann Arbor, MI  48109-1109, USA \\
smoller@umich.edu

\medskip
\noindent
Shing-Tung Yau \\
Harvard University  \\
\noindent
Mathematics Department \\
\noindent
Cambridge, MA 02138, USA\\
yau@math.harvard.edu

\end{document}